# The Third-Party Access Effect: An Overlooked Challenge in Secondary Use of Educational Real-World Data

Hibiki Ito[1*]    Chia-Yu Hsu[1]    Hiroaki Ogata[1]

[1]Kyoto University

*`ito.hibiki.77n@st.kyoto-u.ac.jp`

**Abstract**: Secondary use of growing real-world data (RWD) in education offers significant opportunities for research, yet privacy practices intended to enable third-party access to such RWD are rarely evaluated for their implications for downstream analyses. As a result, potential problems introduced by otherwise standard privacy practices may remain unnoticed. To address this gap, we investigate potential issues arising from common practices by assessing (1) the re-identification risk of fine-grained RWD, (2) how communicating such risks influences learners' privacy behaviour, and (3) the sensitivity of downstream analytical conclusions to resulting changes in the data. We focus on these practices because re-identification risk and stakeholder communication can jointly influence the data shared with third parties. We find that substantial re-identification risk in RWD, when communicated to stakeholders, can induce opt-outs and non-self-disclosure behaviours. Sensitivity analysis demonstrates that these behavioural changes can meaningfully alter the shared data, limiting validity of secondary-use findings. We conceptualise this phenomenon as the *third-party access effect* (3PAE) and discuss implications for trustworthy secondary use of educational RWD.

## 1 INTRODUCTION

The last few decades have witnessed a rapid increase in the use of digital technologies in education, leading to growing real-world data (RWD; Mahajan, 2015) about learning and its environment. Particularly, educational activities that were once transient and confined to the involved learners and educators now leave fine-grained traits of log data that reveal their learning processes (Boulahmel et al., 2025). Educational technology researchers are urging to take advantage of these micro-level data to understand individual learning processes and personalise learning (R. Baker & Siemens, 2014; Romero & Ventura, 2020).

Nevertheless, despite the potential of increasing fine-grained RWD in education, access to such data by third-party researchers has been limited due to privacy constraints (Choi et al., 2020; Fischer et al., 2020), leaving sensitive, but potentially useful data in enclaves (R. Baker & Hutt, 2025). Although protecting the privacy of learners is quintessential in the secondary use of educational data, third-party access is also required for advancing open science and developing reliable educational technologies (R. S. Baker et al., 2024; Fischer et al., 2020).

To ensure an appropriate trade-off between the preservation of individual privacy and the pursuit of collective societal benefits, choosing appropriate privacy protection techniques based on careful risk assessment is of paramount importance (Hartzog & Rubinstein, 2017; Joksimović et al., 2022). Prior research in education domain has investigated privacy risks associated with educational data such as re-identification risk (e.g. Vatsalan et al., 2022; Yacobson et al., 2021) and membership inference risk (e.g. Liu et al., 2024, 2025). To mitigate these risks, privacy-preserving data sharing methods such as k-anonymity (e.g. HarvardX, 2014; Kuzilek et al., 2017), synthetic data generation (e.g. Flanagan et al., 2022; Khalil et al., 2025) and differential privacy (e.g. Gursoy et al., 2017; Ito et al., 2026) have been studied in the context of education. In addition, risk communication with stakeholders is essential for trustworthy data sharing (Slade & Prinsloo, 2013). Prior work explored privacy perceptions and attitudes of stakeholders including students (e.g. Korir et al., 2023; Mutimukwe et al., 2022), parents (Primaningtyas et al., 2025) and instructors (e.g. Francis et al., 2023; Jones et al., 2022). Recent research also investigates effective communication strategies for data sharing decision (Beardsley et al., 2019; Longin et al., 2025; Montenegro et al., 2025).

However, most prior work on secondary use of educational RWD focused on limited scopes in isolation such as "is the data safe to access?" and "do the stakeholders accept?" The purpose of secondary use is not to enable data access, but to generate scientific findings that inform research, system design and educational practice, which eventually benefits stakeholders and the broader population. Thus, these privacy practices must be evaluated by their effects on downstream analyses, that is, how privacy protections and stakeholder communication affect the validity of the findings derived from secondary-use analyses. Nonetheless, it remains largely underexplored how these privacy practices jointly shape the data shared with third parties. As a result, otherwise conventional privacy practices may introduce problems that remain unnoticed.

To fill this critical gap, we investigate problems arising from common privacy practices in secondary use through three analyses. First, we assess the re-identification risk of fine-grained educational RWD anonymised by common methods. We focus on fine-grained trace data because it is a common type of RWD in education and provides rich research opportunities to understand and support learning (Fischer et al., 2020). Second, we explore how communicating the quantified risk while clarifying the secondary-use purpose influences learners' privacy behaviour such as data sharing decisions and self-disclosure behaviours. Although learners are rarely informed about such risk in practice (Beardsley et al., 2019), the absence of risk communication does not eliminate downstream effects; it merely renders them unobservable. Even when researchers do not explicitly communicate re-identification risks, stakeholders may become aware of such risks through external channels such as media coverage of data breaches and public re-identification demonstrations (Sweeney et al., 2018). Finally, we investigate the sensitivity of secondary-use findings to the privacy behaviour associated with third-party access and risk communication. We focus on a learning habits study as a typical example of secondary use of RWD and evaluate the sensitivity of analytical conclusions to plausible increases in opt-outs. Though opting out is not the only factor that affect downstream analyses (e.g. decreased engagement due to privacy concern), we focus on opt-outs because observing non-self-disclosure behaviours requires more costly experiments with retrospective consents, which may not be ethical or feasible in practice (Tassé et al., 2010). Taken together, the present paper addresses the following research questions (RQs):

**RQ1**: What is the re-identification risk of fine-grained educational RWD anonymised by common methods?
**RQ2**: How does communicating quantified re-identification risk influence learners' willingness to share data and self-disclosure behaviours?
**RQ3**: How sensitive are findings from secondary-use analyses of learning habits data to plausible increases in opt-outs from data sharing?

## 2 BACKGROUNDS
This section defines the secondary use of educational RWD and clarifies the technical and socio-technical dimensions that we focus on in this paper.

### 2.1 Secondary use of RWD
Since the early period, it has been acknowledged that educational technology research is inherently constrained by privacy issues (Pardo & Siemens, 2014; Slade & Prinsloo, 2013). Experimenting with a new system in the real world is costly as it requires the involvement of stakeholders and careful protection of personal data (Marshall et al., 2022). Therefore, to complement experimental data, RWD, not primarily collected for research purposes, has gained growing attention (Kuromiya et al., 2023; Mahajan, 2015). Nevertheless, the access to it is largely limited (Choi et al., 2020; Fischer et al., 2020). We refer to secondary use of RWD as sharing personal data collected and stored within an institution with third-party researchers, typically those outside of the institution. In this paper, we use the terms secondary use, data sharing and third-party access interchangeably.

Related recent studies include the attempts to extract real-world evidence (RWE) from RWD (Kuromiya et al., 2023; Okumura et al., 2026). While these works aim to automate the feedback to practitioners based on extracted RWE (i.e. primary use), our goal is to share the data with researchers, so they generate findings by their own methods. Nonetheless, as mentioned previously, the secondary-use findings are expected to eventually benefit the stakeholders and the broader population through research, system development and evidence-driven education practice.

## 2.2 Technical dimension of secondary use

We focus on privacy protection through anonymisation techniques, a core technical issue of secondary use. This problem arises when other security-related technical issues are solved, such as access control and protection against cyber-attacks, which are out of our scope. The protection of private information about individual learners is also a key concern among learners and other stakeholders (Guo et al., 2025; Korir et al., 2023).

Among various privacy risks, we focus on the re-identification risk because it is the most critical risk and the major concern in many privacy policy frameworks such as general data protection regulation (GDPR; Dwork et al., 2017; Gadotti et al., 2024; Salem et al., 2023). Generally, re-identification refers to the identification of an individual whose data is "anonymised" by removing personally identifiable information such as the name and email. Re-identification is typically realised by exploiting auxiliary information from other sources such as online information (Narayanan & Shmatikov, 2008; Sweeney et al., 2018; Yacobson et al., 2021). For example, as shown in Figure 1, combining several social media posts could allow for uniquely determining the student in the dataset. A successful re-identification would leak all student information associated with the same ID, including possibly more sensitive information such as final marks of every module. Notably, even if prohibited by a data sharing policy or regulation, re-identification is undetectable, thereby assessing its possibility in advance being crucial.

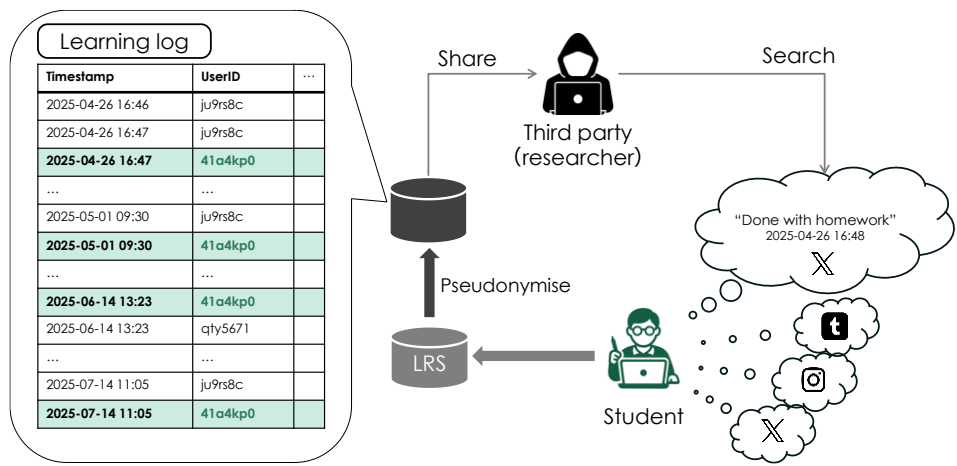

Figure 1: Example workflow of re-identification by a third-party researcher exploiting public information on social media. Learning log data is stored in a learning record store (LRS) and shared with a third-party researcher after pseudonymisation. The researcher re-identifies a target student by linking information from the student's social media posts.

In education domain, the ARX (Prasser et al., 2014; Prasser & Kohlmayer, 2015) is perhaps the most used assessment tool of re-identification risk of educational data. Another tool for assessing re-identification risk is the Re-identifier Risk Ready Reckoner (R4) developed at the Commonwealth Scientific and Industrial Research Organisation (CSIRO, 2019). Unlike the algorithm of ARX, it quantifies re-identification risk using Markov models accounting for not only uniqueness but also uniformity, i.e., consistency of individuals throughout the course (Vatsalan et al., 2022).

Nonetheless, both these tools take as input tabular data and are not applicable to set-valued data. Particularly, little attention has been paid to risk assessment of fine-grain educational data, despite the prevalence as RWD.

Proposed by de Montjoye et al. (2013), unicity is a framework to evaluate the re-identification risk based on the uniqueness of individuals in a set-valued dataset and has been applied to mobility trace data (de Montjoye et al., 2013; Farzanehfar et al., 2021; Kondor et al., 2020; Pellungrini et al., 2018; Xu et al., 2018), transaction data of credit cards (de Montjoye et al., 2015), web browsing histories (Deuser et al., 2020) and installation histories of smart phone applications (Achara et al., 2015; Sekara et al., 2021). It should be noted that re-identifiability and uniqueness are distinct concepts, and unicity is a metric of uniqueness, potentially overestimating re-identification risk (Barth-Jones et al., 2015). However, unicity provides a realistic measure of how much auxiliary information would suffice for re-identification, as uniqueness immediately implies identification, and the availability of such auxiliary information should not be underestimated in the era of artificial intelligence (Carlini et al., 2025).

## 2.3 Socio-technical dimension of secondary use

The socio-technical dimension of secondary use consists of complex dynamic actions at different layers such as national and organisational levels (Baxter et al., 2023). Among them, we focus on communication with stakeholders, especially learners, because it directly affects the downstream secondary-use analyses by e.g. their opt-outs to data sharing. Many policy frameworks and guidelines on sharing of educational data emphasise transparency and appropriate communication with stakeholders (Drachsler et al., 2010; Greller & Drachsler, 2012; Khalil & Ebner, 2016; Pardo & Siemens, 2014; Tsai & Gasevic, 2017). There has also been empirical evidence that learners and other stakeholders value the transparency when their data is shared with third parties (Cerratto Pargman & McGrath, 2021; Korir et al., 2023; Mutimukwe et al., 2022; West et al., 2020; Yang & Hasan, 2023).

While most empirical work focused on perceptions of stakeholders about third-party access (Cerratto Pargman & McGrath, 2021), little is known about how communication strategies impact their data sharing decisions and non-self-disclosure behaviour. Concurrent studies have shown that learners' willingness to share data varies across contexts and tends to decrease for secondary use, particularly when communication involves group discussions (Longin et al., 2025; Montenegro et al., 2025). However, these studies typically assume anonymised data and do not communicate re-identification risk, leaving a gap between socio-technical research on data sharing and technical privacy risk assessment. Moreover, despite opt-in, a learner could exhibit non-self-disclosure behaviour such as intentional disengagement with the system at certain times when they are informed about the risk. This is related to chilling effects in broader data privacy studies (Stubenvoll & Binder, 2024) and observer effects in behavioural sciences (Baclawski, 2018; Saha et al., 2024). A notable example of the former includes the significant drop in traffic on privacy-sensitive Wikipedia articles after revelation of US government's massive online surveillance (Penney, 2016). The latter refers to behavioural changes in research participants due to the awareness of being observed by researchers. A remarkable difference between these prior works and non-self-disclosure behaviour induced by third-party access is that, while chilling and observer effects assume that an observer knows the identity of subjects, secondary use is typically conducted by anonymised, at least pseudonymised, data, introducing uncertainty in identifiability. In other words, being observed remains a risk, rather than a fact. Nevertheless, it is unknown whether learners show such non-self-disclosure behaviour when aware of re-identification risk.

## 3 METHODS

We conduct three analyses to address each RQ: risk-assessment, a vignette survey and sensitivity analysis. We first describe data and contexts for these analyses and then delineate analytical methods.

### 3.1 Data and contexts

For RQ1 and RQ3, we use the same datasets (see Table 2): two institutional datasets collected through BookRoll, an e-book system (Ogata et al., 2017), Duolingo lesson logs (Settles, 2017) and the EdNet-KT4 dataset (Choi et al., 2020). BookRoll allows for distributing learning materials as PDFs. As shown in Figure 2, interaction logs (e.g. open/close materials) are sent to the learning record store (LRS) in the form of the xAPI standard. Each log includes the timestamp, actor, xAPI ver, etc. Additionally, each xAPI log is associated with a BookRoll-specific operation name, which indicates more granular description than xAPI verbs. Table 2 shows some examples of operation names and corresponding xAPI verbs.

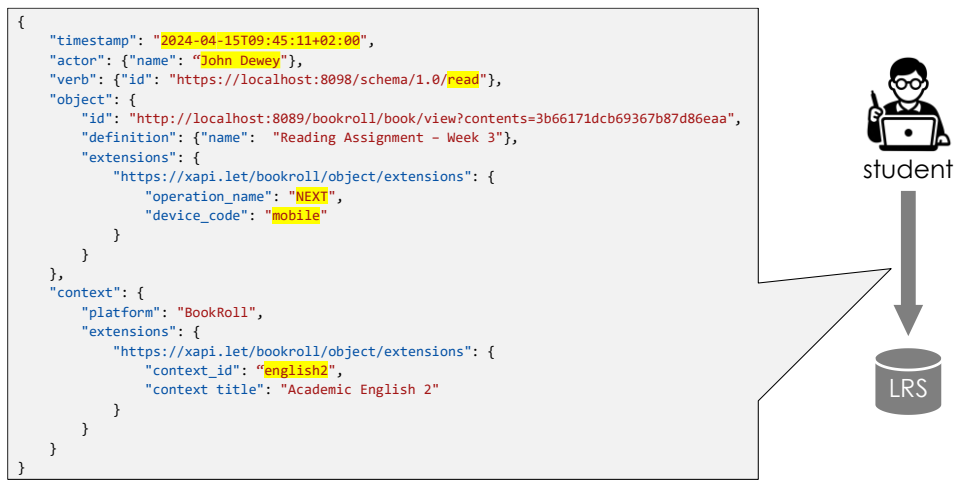

Figure 2: Example of xAPI log collected by BookRoll.

Table 1: Examples of BookRoll-specific operations and xAPI verbs.

| xAPI verb | Operation Name | Function |
|---|---|---|
| read | NEXT | Go forward to the next page |
|  | PREV | Go back to the previous page |
| noted | ADD_MEMO | Add a memo |
|  | DELETE_MEMO | Delete a memo |
| highlighted | ADD_MARKER | Add a marker highlight |
|  | DELETE_MARKER | Delete a marker highlight |

We employ two BookRoll datasets obtained at a public university and a secondary school. The context of the BookRoll University dataset is specific to a single module with medium class size (n=51), and the duration of data collection ranges over a semester, thus being a common data unit for education research. On the other hand, the BookRoll Secondary dataset consists of Japanese secondary school students' log data (n=752) across multiple classes (mostly mathematics and English). The scope of the data is limited to a single school, which is a typical unit of educational data in K-12 settings for primary use. As the dataset is larger than BookRoll University in terms of volume and represents more contextual diversity, it would be suitable for studying cross-context learning behaviour in a secondary-school setting. For fair comparison, we set the period of data collection to a single semester, the same as for the BookRoll University dataset. This serves as a lower bound of more longitudinal data since the uniqueness of individual trajectories increases as data becomes more longitudinal.

The Duolingo and EdNet datasets help generalising and deepening insights on BookRoll data. The former is open data for replication of a study on second language learning by Settles and Meeder (2016). It contains Duolingo vocabulary lesson logs for each learner, where each record includes the timestamp, lexeme (i.e. the target word) and learning language. Although the dataset spans only a two-week period, it reflects more diverse self-paced learning behaviours, as learners engage with Duolingo lessons independently—unlike the BookRoll data, where learner activity is shaped by regular, scheduled classes in formal educational settings. The EdNet (more precisely, EdNet-KT4) dataset is a large-scale open dataset consisting of exercise results on Santa, a multi-platform self-study application for preparation of the TOEIC (Test of English for International Communication) exam (Choi et al., 2020). The data provides more longitudinal learning processes of a larger number of learners, complementing the Duolingo dataset. The results derived from these public datasets also exemplifies a common form of educational data that are gathered and stored by educational technology companies, yet such RWD often remains inaccessible to researchers for secondary use. Our findings could serve as evidence to encourage companies to adopt effective privacy protection methods for data sharing, which is essential for advancing the broader use of educational data (Fischer et al., 2020).

Table 2: Summary of datasets.

| Dataset | # records | # students | Period | Description |
| --- | --- | --- | --- | --- |
| BookRoll University | 66 259 | 51 | 4 months | Reading behaviour in a bachelor-level academic reading module |
| BookRoll Secondary | 6 486 986 | 752 | 4 months | Reading behaviour in multiple secondary school classes (mostly mathematics and English) |
| Duolingo | 12 854 226 | 115 222 | 2 weeks | Duolingo vocabulary lessons for multiple learning languages |
| EdNet-KT4 | 131 441 538 | 297 915 | 15 months | English reading and listening exercises |

The participants of the vignette survey for RQ2 (see Section 3.3 for the method) are undergraduate students at a Japanese public university (n=19). The limited sample size does not pose a significant problem to our analysis because our goal is not to discover nomothetic, population-level behavioural patterns but rather to explore potential problems. Since benefits of secondary use can be both collective and individualistic, learners' data sharing decision or non-self-disclosure behaviours may unfairly cause detrimental downstream effects on themselves due to third-party access. For example, the opt-out to contributing to train a cross-institutional centralised machine learning model that supports personalised learning could bias predictions for the opt-out learners, and, even worse, the use of this system could amplify this negative feedback over time (Archer & Prinsloo, 2020). From this perspective, even with a few samples, influence of risk communication on data sharing and self-disclosure behaviour can be an important challenge that needs to be studied. Unlike typical survey studies where vignettes are unrelated to participants' real life, the participants of the study use BookRoll for their own university studies, and the actual re-identification risk of BookRoll data identified in our risk assessment is used in the vignettes. This way, the vignettes very closely reflect the real life of participants, except that the participants' data is not really shared with third parties.

### 3.2 Analysis 1: Risk assessment (RQ1)

We apply the unicity framework (de Montjoye et al., 2013) to the four datasets[1]. Algorithm 1 shows the process of calculating the unicity, given a dataset and a set of quasi-identifiers (QIs). Roughly, unicity indicates how many events suffice to single out a student in a dataset. We estimate the unicity by taking the average of the outputs over ten random seeds. For the BookRoll University and Secondary datasets, we include the entire datasets in each sample (i.e. the sample size $m$ equals the number of students in each dataset). For the Duolingo and EdNet datasets, we set $m = 2500$.

In our experiments, we focus on timestamps as a main QI, as they are common in set-valued log data and also play an important role in temporal analyses of learning processes (Knight et al., 2017), thereby being a key to balance privacy and utility. We assume that the attacker knows that the target student is in the dataset. While the raw data contains timestamps recorded to seconds, it is not very realistic for an adversary to infer learner behaviours at second-

---

[1] See the supplementary material for the code at https://anonymous.4open.science/r/3PAE.

**ALGORITHM 1:** CALCULATE UNICITY

**Input**: $\mathcal{D}$ (dataset), $Q$ (set of QIs), $\varepsilon$ (number of events available to the attacker), $m$ (sample size)
**Output**: *Unicity*

1.   $D \leftarrow$ subset $\mathcal{D}$ by quasi-identifiers in $Q$            // only quasi-identifiers are used
2.   $S \leftarrow$ randomly choose $m$ students from $D$
3.   *Unicity* $\leftarrow 0$
4.   **for** $t$ in $S$ **do**            // go through every student in the sample
5.      *Trajectory* $\leftarrow$ all events of $t$ in $D$
6.      *Observations* $\leftarrow$ randomly choose $\varepsilon$ events in *Trajectory*
7.      *Candidates* $\leftarrow$ choose from $D$ students whose events include *Observations*
8.      **if** |*Candidates*| $= 1$ **do**            // check if the target is unique
9.           *Unicity* $\leftarrow$ *Unicity* $+1$
10.      **end if**
11.   **end for**
12.   **return** *Unicity* $/ m$

level precision from auxiliary information. Therefore, we pre-process all the datasets by converting timestamps into minute-level precision and use them as a QI. It is assumed that this level of temporal precision should not significantly affect downstream analyses.

While pseudonymisation is a common approach to privacy protection when sharing educational data (Amo et al., 2020; Hutt et al., 2023), we also consider naïve coercing (a.k.a. generalisation) as an example anonymisation method because it is an established technique for set-valued data (Samarati & Sweeney, 1998), easy to automate (Prasser et al., 2014) and particularly favoured by practitioners as a safe choice due to the non-perturbative nature (Guo et al., 2025). The unit of timestamps are generalised to longer time windows such as quarters, hours and dates (see Figure 3). Consequently, while everyone becomes less unique, we may relax the assumption on attackers such that they only have rough observation that a target student is active on the learning platform within a time window. For example, if there is only one student whose (possibly multiple) BookRoll logs fall in a certain time window, the student can be re-identified by observing that the student uses BookRoll within that time window.

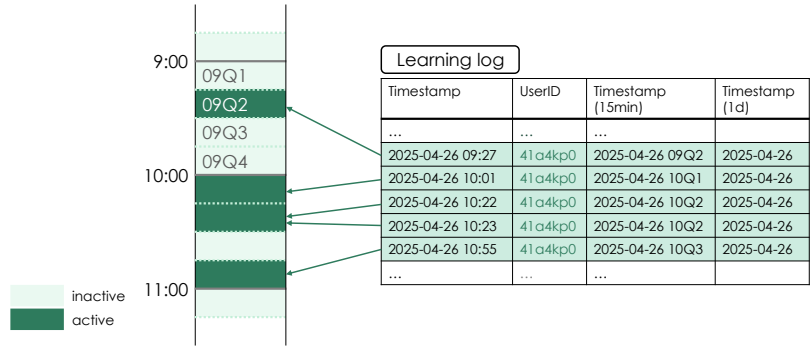

Figure 3: Generalisation of timestamps.

## 3.3 Analysis 2: Vignette survey (RQ2)

Since experimenting risk communication and observing data sharing and self-disclosure behaviours is costly and potentially unethical, we instead perform a vignette survey. The survey consists of three steps: first, the participants are asked if they would give a consent to sharing their non-anonymised raw data on a digital learning platform such as BookRoll with third-party researchers. Second, pseudonymisation is explained with an example, and the participants are asked if they would share their pseudonymous data with third-party researchers. Finally, the quantified re-identification risk is explained using the above-mentioned example where an attacker exploits social media posts, and then they are asked if they would share their pseudonymous data with third-party researchers.[2] By presenting comparable scenarios, the survey is designed to better capture participants' actual behaviours (Hainmueller et al., 2015)

For each of the three scenarios, we asked about digital learning environments in general, using BookRoll as an example. We clearly explained that the data is shared only for research purposes. For each of the three questions, those who answered "Yes" (i.e. willing to share) are asked an optional open-ended question on how data sharing would affect their use of the system and social media. Though social media posts are just an example of auxiliary information that an attacker can exploit, it is a realistic re-identification attack model and posting learning progresses on social media could also influence learning processes (Klein et al., 2020), thereby being the focused example in the study. This survey study is approved by the institutional review board of the university that the participants belong to.

### 3.4 Analysis 3: Sensitivity analysis (RQ3)

The approach of sensitivity analysis is employed to evaluate consequences of the risk communication for downstream secondary-use analyses. That is, we study the sensitivity of conclusions of secondary-use analyses to plausible increases in opt-outs. Though non-self-disclosure behaviours despite opt-in could also affect secondary use, it is hard to estimate such effects solely from the vignette survey, leaving empirical examination to future work. We focus on learning habits research as an example secondary use of educational RWD. The reason is twofold: there has been evidence that forming a habit of learning—defined as a repetitive behaviour in the context of learning (Wood & Neal, 2007)—has a significant effect on learning such as academic achievement (Boroujeni et al., 2016; Ricker et al., 2020) and productivity (Hsu et al., 2024); and, since learning habits research typically uses system-agnostic indicators such as time-on-task and time of day, we are able to set the same variables across the four datasets. We focus on ordinary least squares (OLS) linear regression that predicts the strength of learning habits from relevant features, a basic example of statistical exploration in secondary use of RWD. The simplicity of linear regression allows us to estimate the worst-case influence of increased opt-outs on its findings.

First, we simulate feature engineering and model selection of third-party researchers by the approach of automated data science (AutoDS). It uses large language models (LLMs) to automate a series of data science tasks, expanding the traditional automated machine learning (AutoML) approach (Hollmann et al., 2023; Li et al., 2025). We employ the AutoDS approach instead of real human analyses to cover a wider range of analytical procedures. Since our datasets contain private information, we only give LLMs the context, data schema and a description of the outcome variable and ask to write a Python script that extracts independent variables avoiding coupling with the dependent variable and perform model selection. Specifically, we only use timestamps for generating features. After extracting features, OLS linear regression is performed with the dependent variable defined as the entropy of time spent on the system over the four time windows, following Hsu et al. (2023): morning (00:00-04:59), afternoon (05:00-11:59), evening (12:00-16:59) and overnight (17:00-23:59). This is similar to the regularity metric proposed by Boroujeni et al. (2016) which calculates entropies over 24 hour bins. Since there are four time windows, the habit entropy falls between 0 and 2, and the smaller the entropy is, the stronger a student has a habit of learning on certain time windows. Models with up to five features that explain the outcome the best are selected. For each dataset, we give similar prompts to `Gemini 2.5 Pro`, `GPT5` and `Sonnet 4` by changing the dataset description and repeat this three times.[3]

The sensitivity analysis on each linear regression model is performed by the approximate maximum influence perturbation (AMIP) metric introduced by Broderick et al. (2020). This allows us to estimate how many opt-outs would be sufficient to change in the worst case the significance, sign or both for each coefficient of an OLS linear regression model (Broderick et al., 2020). We use the authors' implementation in the `zaminfluence` package in R[4].

---

[2] See the supplementary material for the vignettes and questions at https://anonymous.4open.science/r/3PAE.

[3] See the supplementary material for all prompts and generated scripts at https://anonymous.4open.science/r/3PAE.

[4] https://github.com/rgiordan/zaminfluence

# 4 RESULTS

## 4.1 RQ1: Re-identification risk of fine-grained RWD

### 4.1.1 Pseudonymisation

Figure 4 shows the estimated unicity for each dataset with 95% confidence intervals (CI) computed by bootstrapping (the subsequent plots also show CIs in the same way). Here, $\varepsilon$ denotes the number of timestamps that the attacker knows belong to the target student based on some auxiliary information, and the timestamps are the only QI. That is, an attacker only knows whether a target student is active on the system at certain time points (i.e. one-minute time windows). For example, the unicity of the BookRoll Secondary dataset for $\varepsilon = 4$ is 0.913, meaning that given four time points an attacker can determine on average 91.3% of the students in the dataset. For smaller $\varepsilon$, the BookRoll Secondary exhibits larger unicity than the BookRoll University, while they converge to the high unicity as $\varepsilon$ increases. This is perhaps because the BookRoll University dataset reflects less diversity by focusing on a specific context, where students regularly attend lectures at the same time, reducing the uniqueness of individual trajectories. Nonetheless, both BookRoll datasets exhibit high unicity with a few observations available to the attacker, implying that pseudonymous log data of learning behaviour is highly susceptible to re-identification.

In addition, despite learners' potential behavioural diversity due to the nature of self-paced learning, the Duolingo and EdNet datasets exhibit lower unicity compared to the BookRoll datasets, converging below 0.6 and 0.3, respectively. This is probably due to the larger numbers of individuals in these datasets. Nonetheless, these values must be interpreted with caution, as the unicity metric only evaluates the average-case re-identification risk. In other words, the unicity framework quantifies the risk averaged over all students, without capturing the variability in individual vulnerability—some students face higher re-identification risks than others. When protecting individual privacy, we are typically interested in the worst-case vulnerability (Ito, 2024). From this perspective, even though the unicity of the Duolingo and EdNet datasets are relatively small, sharing these datasets would still require stronger privacy protection measures.

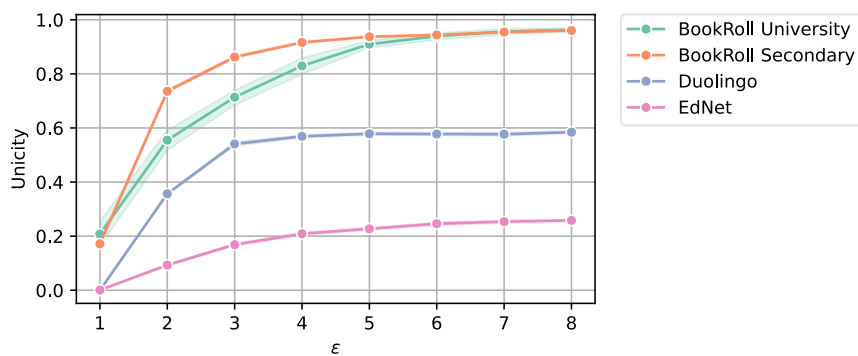

Figure 4: Unicity of each dataset with one-minute timestamps. $\varepsilon$ denotes the number of timestamps which the attacker knows belongs to the target student.

### 4.1.2 Naïve anonymisation

Figure 5 illustrates the changes in unicity of each dataset when generalised timestamps with different levels (one-minute, quarter and date) are used as a QI. Overall, the effectiveness of naïve coercing depends on the nature of a target dataset. For the BookRoll University dataset, several students are not protected from re-identification with a few timestamps even if the timestamps are generalised to dates. This is probably because students do not engage with the learning materials every day, increasing the uniqueness of dates when each student is active on BookRoll. On the other hand, for the other three datasets, the generalisation of timestamps to dates protects almost all learners from re-identifying by at least up to eight dates, implying that this naïve coercing effectively mitigates the re-identification risk (i.e. unicity near zero).

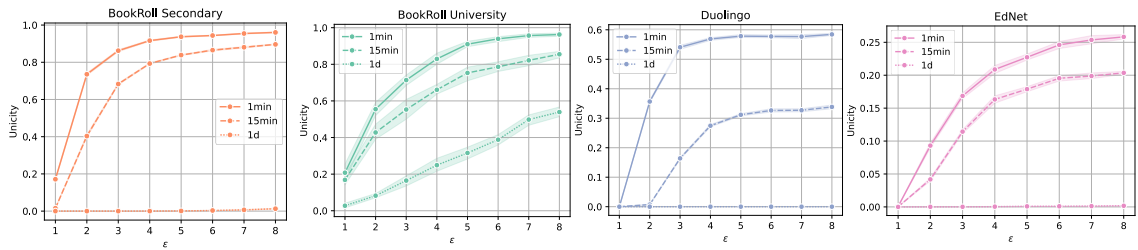

Figure 5: Unicity of different datasets with timestamp generalisation.

### 4.2 RQ2: Influence of quantified risks on data sharing decisions and self-disclosure behaviour

Figure 6 shows the responses to the questions of whether students give permission to share data with third parties for each step. First, when the raw data is shared, most students give the consent. However, the follow-up open question revealed two different patterns of non-self-disclosure behaviours: some students (n=2) stated that third-party access to their raw data would "not influence" their use of the system, while others (n=5) reported their intention to change their behaviour, such as "I would keep the material open so I look studying it" and "note-taking and highlighting would decrease because I don't want to show off my lack of knowledge to others".

On the other hand, after pseudonymisation was explained, no participants reported potential non-self-disclosure behaviours for sharing pseudonymous data. As pseudonymisation is very common in practice (Hutt et al., 2023), our result suggests that the reliability of existing empirical research may not be much affected by learners' non-self-disclosure behaviour.

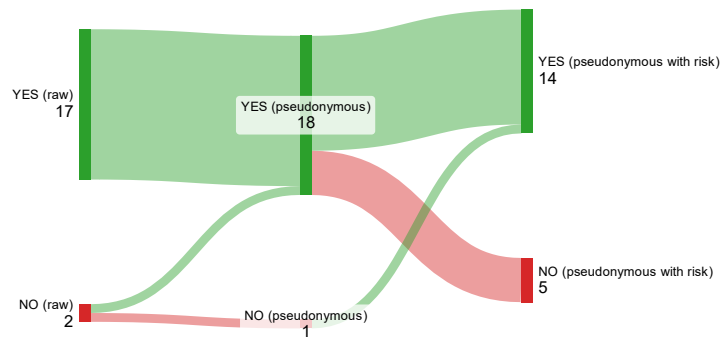

Figure 6: Learners' willingness to share data with third-party researchers. Participants (n=19) were first asked if they would share raw data (left). Next, pseudonymisation was explained, and the participants were then asked if they would share pseudonymous data (centre). Finally, the quantified re-identification risk of pseudonymous data was explained, and they were then asked if they would share pseudonymous data (right).

Nevertheless, a pivotal concern arises when the quantified risk of pseudonymisation is communicated with the participants: none of the participants were aware of the re-identification risk of pseudonymised log data, and some reconsidered their data sharing decisions negatively (n=5). In addition, even among those who answered positively to giving consents, some reported intentions to systematically change their use of social media (n=4). For instance, a student who answered to opt in after the risk is explained wrote: "I would no longer make a social media post that indicates my use of the system." Although changed usage of social media does not directly imply behavioural data on

### 4.3 RQ3: Sensitivity of secondary-use findings to increased opt-outs

Estimating AMIP sometimes fails when the algorithm is not able to find a small proportion of points to change the coefficients (Broderick et al., 2020). Table 3 shows the success rates for each dataset. Compared to the Duolingo and EdNet datasets, the success rates are lower for the smaller BookRoll datasets. This is because the asymptotic linear approximation of AMIP becomes unstable for small data (Broderick et al., 2020). Thus, subsequent results for BookRoll datasets should be seen as rougher approximates than the other datasets.

Table 3: Success rates of changing conclusions.

| Dataset | Significance | Sign | Both |
|---|---|---|---|
| BookRoll University | .85 | .44 | .00 |
| BookRoll Secondary | .78 | .65 | .45 |
| Duolingo | .98 | .96 | .93 |
| EdNet (KT4) | .98 | .96 | .93 |

Within successful AMIP estimations, Figure 7 shows the distributions of proportions needed to remove for flipping the inference conclusions. Across datasets, a substantial share of coefficients exhibits low robustness, with changes in sign or statistical significance occurring after the removal of fewer than 10% of students. When considered alongside Analysis 2—where a nontrivial portion of students reported an intention to opt-out when informed about the re-identification risk—these results indicate that findings from secondary-use analyses can be highly sensitive to plausible increases in opt-outs. For the BookRoll Secondary, Duolingo and EdNet datasets, some coefficients are susceptible to both significance and sign changes under relatively small, non-random data dropout. Such sensitivity suggests that secondary-use findings based on shared data may be unstable under realistic participation scenarios.

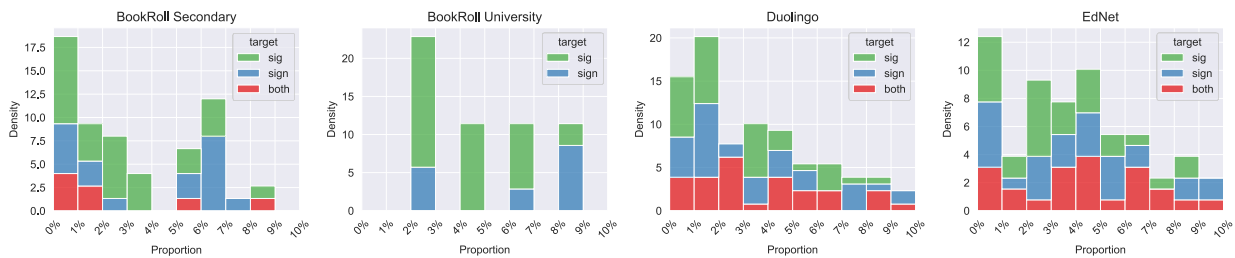

Figure 7: Proportions of students that need to opt-out to change the significance ("sig"), sign and both for coefficients in OLS linear regression models predicting the habit strength.

It should be noted that, even if the sign or significance of a coefficient changes by dropping some data points, it does not mean that the findings on the shared data are wrong. Rather, it suggests that results are true for the sample but unlikely to generalise to the population or different groups of learners. Because opt-outs are rarely random (Marland et al., 2020), it is generally not possible to infer from the shared data alone what conclusion would have been obtained without such opt-outs. Consequently, secondary-use analyses with limited robustness are particularly vulnerable to participation changes associated with common privacy protection techniques and their communication. This highlights the need for stronger privacy-preserving methods and more effective stakeholder communication strategies to support the validity of downstream secondary-use research.

# 5 DISCUSSIONS

## 5.1 Key findings: third-party access effect (3PAE)

Our results indicate that common anonymisation and stakeholder communication practices can introduce nontrivial validity risks for downstream secondary-use analyses conducted by third-party researchers. Specifically, we showed that fine-grained educational RWD entails high risk of re-identification when merely pseudonymised or naively anonymised (RQ1), that communicating such quantified risks nontrivially affects learners' intention to share their data with third parties and potential non-self-disclosure behaviours on the system (RQ2), and that downstream secondary-use analyses can be highly sensitive to resulting changes in the shared data (RQ3). Because these findings arise from a complex chain of technical and socio-technical processes, it is beneficial to conceptualise such phenomenon explicitly in order to discuss implications and guide future research (Gerring, 1999; Podsakoff et al., 2016). This is particularly important for trustworthy secondary use of educational RWD, where the concept of privacy—a core component of trustworthiness—is essentially contested (Mulligan et al., 2016) and the term has been used very ambiguously or inconsistently in the educational technology literature (Viberg et al., 2022).

Based on these considerations, we define third-party access effect (3PAE) as *a phenomenon that what third parties learn from data differs from what would otherwise be the case without such third-party access*. Our results should not be interpreted as evidence that 3PAE is realised in all secondary-use settings, but instead as demonstrating the risk of 3PAE arising under common, plausible privacy practices and learner responses. In the following, we justify the definition by using the eight criteria of good concept formulation by Gerring (1999):

1. Familiarity: 3PAE builds on well-established conceptual lineages such as chilling effects and observer effects.
2. Resonance: The term "third-party access" foregrounds the institutional and procedural conditions that trigger the effect, rather than relying on the ambiguous notion of privacy alone.
3. Parsimony: Naming of 3PAE works as an effective abbreviation of the intricated phenomenon, compressing a complex causal chain of privacy protection, stakeholder communication, learner behaviour and biassed downstream analyses.
4. Coherence: The definition is explicitly structured as a counterfactual comparison of third-party access, a clear causal condition.
5. Differentiation: 3PAE is neither an observer effect nor a chilling effect, both of which involve non-anonymised data, nor is it merely selection bias as understood in the general statistics literature, which lacks institutional privacy triggers.
6. Depth: It integrates multiple aspects of trustworthy secondary use that have previously been studied in isolation by identifying their shared downstream effect on what third parties can learn from data.

To address the remaining criteria—theoretical utility and field utility—we situate 3PAE in relation to differential privacy (DP), a well-established formal notion of privacy (Dwork et al., 2006). Although DP originates from computer science, its central intuition is simple: conclusions drawn from data analysis should not meaningfully depend on the participation or behaviour of any single individual. Viewed through this lens, 3PAE captures, at the level of human behaviour and inference, the same sensitivity problem that DP formalises at the level of datasets and algorithms: nonparticipation corresponds to who is included in the dataset, while non-self-disclosure corresponds to non-random differences between observed and counterfactual behaviours.

7. Theoretical utility: This conceptual connection provides theoretical grounding for 3PAE by linking it to a well-established privacy framework.
8. Field utility: Theoretically grounded while applicable beyond technical implementation, 3PAE offers a common conceptual frame for communicating between the technical and socio-technical research communities towards trustworthy secondary use.

## 5.2 Implications

Our findings indicate that common privacy practices in the secondary use of educational RWD can give rise to conditions under which 3PAE may pose a validity risk for downstream analyses. The risk of nonignorable 3PAE underscores the need for developing improved methodologies and practices to mitigate such risks. Since merely pseudonymised or naively anonymised trace data poses high re-identification risks, advanced anonymisation techniques should be tailored to educational RWD to achieve privacy by design (Cavoukian, 2009). Though research on anonymising educational RWD is growing, prior work predominantly focuses on statistical trade-offs in isolation, with limited attention to socio-technical dynamics. From the perspective of 3PAE, evaluations of privacy protection techniques should therefore consider not only privacy-utility trade-offs, but also their potential implications for downstream secondary use through stakeholder communication. Our results highlight the importance of studying risk communication practices alongside technical protections, as communicated risk assessments can plausibly influence participation and self-disclosure behaviour. While the rich literature of risk communication could provide foundations in stakeholder communication (Aven, 2018), the lens of 3PAE offers a way to develop communication strategies specifically tailored to secondary use of RWD in education, with explicit attention to downstream analyses.

As discussed above, 3PAE is conceptually connected to DP. As such, among other anonymisation methods, implementing and appropriately communicating DP has the potential to directly mitigate both channels of 3PAE. Under DP, the presence, absence, or behavioural variation of any single individual has a bounded influence on downstream outputs, rendering both nonparticipation and strategic behaviour effectively meaningless at the individual level. Although DP has been used to anonymise educational data, its socio-technical issues remain unknown in the context of education. Particularly, while communicating DP with stakeholders, typically not aware of DP, is an emerging challenge in the privacy research community (e.g. Bullek et al., 2017; Nanayakkara et al., 2023; Xiong et al., 2023), studies focusing on secondary use of RWD are scarce. In this regard, 3PAE provides a useful conceptual framework for guiding future research on DP in education across both technical and socio-technical dimensions, contributing to more trustworthy secondary use of educational RWD and to broader discussions in privacy research.

While 3PAE is not limited to secondary use of educational data, it is particularly relevant to the education domain because a major part of RWD in education consist of interaction logs derived from digital learning environments, which can be affected by non-self-disclosure behaviours associated with third-party access within the learning environment. This contrasts with, say, medical RWD such as data collected through wearable devices, where heart rates or sleep patterns are unlikely to be influenced by third-party access.

## 5.3 Limitations and future work

Although unicity considers a realistic threat model and provides evidence of high re-identification risk, it is not the only metric for assessing re-identification. Other possible forms of re-identification of fine-grained educational data would need further investigation. In addition, behavioural validity of the vignette study remains untested, despite the strategic design to reflect participants' real-world contexts. Moreover, our sensitivity analysis may not fully capture what researchers analyse in practice. While this study explores potential problems in common privacy practices, experimentally validating the phenomenon of 3PAE is left to future work.

The theoretical connection between 3PAE and DP remains conceptual in this work. As discussed in the previous section, the combination of technical and socio-technical studies of DP for secondary use of educational RWD is an important future direction towards mitigating 3PAE. Though DP is not the only solution, it serves as the gold-standard from both technical and socio-technical perspectives through the lens of 3PAE. Further empirical and theoretical investigation of 3PAE—such as connection to membership inference (Niu et al., 2024)—would be a potential area of future work. Such work would contribute to a more comprehensive understanding of the conditions under which 3PAE poses meaningful risks and how these risks can be mitigated in the secondary use of educational RWD.

# 6 CONCLUSION

To support trustworthy secondary use of educational RWD, investigation on privacy protection techniques and stakeholder communication is essential, yet predominant prior work focused on either technical or socio-technical dimensions in isolation. As a result, potential issues arising from common privacy practices remain unknown. Thus, this paper examined how common privacy practices jointly affect downstream analyses. We showed that high risk of re-identification, through naïve communication with stakeholders, can increase opt-outs and non-self-disclosure intentions, and that downstream analytical conclusions are highly sensitive to these plausible changes. Together, these findings highlight a mechanism through which privacy practices and their communication may introduce validity risks for secondary-use analyses, when standard privacy practices are applied. The concept of 3PAE, highlighted by these findings, offers a unifying conceptual lens to bridge the technical and socio-technical dimensions of privacy studies, contributing to ongoing efforts to advance more trustworthy secondary use of RWD in education.